  \documentstyle[12pt,procsla]{article}
  
  \catcode`\@=11
  \long\def\@makefntext#1{
  \protect\noindent \hbox to 3.2pt {\hskip-.9pt  
  $^{{\ninerm\@thefnmark}}$\hfil}#1\hfill}		
  
  \def\@makefnmark{\hbox to 0pt{$^{\@thefnmark}$\hss}}  
  	
  \def\ps@myheadings{\let\@mkboth\@gobbletwo
  \def\@oddhead{\hbox{}
  \rightmark\hfil\ninerm\thepage}   
  \def\@oddfoot{}\def\@evenhead{\ninerm\thepage\hfil
  \leftmark\hbox{}}\def\@evenfoot{}
  \def\sectionmark##1{}\def\subsectionmark##1{}}
  
\begin{document}
  
  \centerline{\normalsize\bf Is the Universe Transparent to TeV Photons?}
  \centerline{\footnotesize W. Klu\'zniak}
  \baselineskip=13pt
  \centerline{\footnotesize\it Physics Department, University of Wisconsin,
   1150 University Ave.,}
  \baselineskip=12pt
  \centerline{\footnotesize\it  Madison, WI 53706, USA}
  \centerline{\footnotesize E-mail: wlodek@cow.physics.wisc.edu}
  

  \abstracts{Direct observations with ground-based instruments have
  shown some relatively nearby ($z=0.03$) extragalactic objects
  to be powerful sources of electromagnetic radiation in the TeV
  range.  It is thought that such radiation cannot travel
  much farther through the intergalactic space, but the actual degree to
  which TeV photons from the (as yet undetected) more distant sources
  are absorbed by the intervening infra-red background is still an
  open question, and one which will certainly be resolved by continued
  monitoring of the sky with present and future detectors. If, as has
  recently been suggested in the context of attempts to quantize
  gravity, Lorentz invariance is broken at an energy scale, $E_Q$
  (greater than the GUTS scale), the kinematics of photon-photon
  collisions would be profoundly affected at much lower
  energies. Specifically, electron-positron pair creation on soft
  photons may be forbidden at energies as low as $30\,{\rm
  TeV}\sqrt{E_Q/10^{17}\,{\rm GeV}}$ and the Universe would then be
  transparent to high energy photons. The proposition that Lorentz
  invariance is broken may be verifiable by the techniques of TeV
  astronomy. 
  }
   
  \normalsize\baselineskip=15pt
  \setcounter{footnote}{0}
  \renewcommand{\thefootnote}{\alph{footnote}}

  \section{Broken Lorentz invariance?}

Lorentz invariance has, of course, been found to be
satisfied to a high degree in all
experiments performed to date. However, it has long been realized that
attempts at finding a quantized description of space-time may lead to
the appearance of non-Lorentz invariant terms. For example, applying
quantum deformations to the 4-dimensional Poincar\'e group results in
a so called $\kappa-$deformed Poincar\'e algebra$^1$, which corresponds to
discretizing time, while ``preserving almost all classical properties
of three-dimensional euclidean space.'' The  $\kappa-$deformation
leads to a distorted mass shell condition
of the form$^{1,2}$ $m^2+{\bf p}^2=[2\kappa\sinh (p_0/2\kappa)]^2$,
with similar changes to the law of energy-momentum
conservation. Here, $\kappa$ is presumably the energy scale at which
Lorentz invariance no longer holds accurately.

More recently, it has been suggested$^3$ that in a wider class of approaches to
quantizing gravity the photon dispersion relation would be
$$pc= E\sqrt{1 + E/E_{QG}},\eqno(1)$$
where $E_{QG}$ could be as low as
$E_{QG}\ge10^{16}\,$GeV$\sim10^{-3}E_{Planck}$, and would presumably be
related, e.g., to some characteristic length scale on which the discrete
nature of space-time becomes apparent, $l\sim E_Q^{-1}$. One consequence$^3$
of  photons having a three-momentum of the magnitude given by
eq. (1), would be that high-energy electromagnetic radiation would travel
with a speed dependent on the photon energy,
$v=c(1-E/E_{QG})$, where $c$ is the ordinary speed of light.

In the discussion below, I assume that the photon dispersion
relation of eq.  (1) is valid, and show,$^4$ that it would have drastic effects
on the kinematics of photon-photon collisions. When the soft photon
has energy $E_1$ in the optical or infrared range, these new effects
would appear when the energy, $E_2$, of the other photon is in the TeV range,
more precisely, when $E_2\sim\sqrt{E_1E_{QG}}$. This would make the
currently observable Universe transparent to high energy
photons. Similar considerations,$^{5,6}$ suggest that the
Greisen-Kuzmin-Zatsepin cutoff would also not hold, i.e.  the universe
would be transparent to multi-PeV protons, as well.

  \section{Extragalactic TeV astronomy and the IR background}

TeV photon astronomy is a well established field with the spectrum of
at least one steady Galactic source (the Crab pulsar) reliably and
reproducibly determined,$^7$ through observations of the {\v C}erenkov
radiation of atmospheric showers. Numerous powerful extragalactic
sources, as well, are expected to exist in the violent Universe.
However, only the closest are thought to be observable, because of
severe attenuation$^{8,9}$ of TeV radiation over distances much larger
than $\sim100\,$Mpc by pair creation on the infrared background (IR), and
(over much shorter distances) of $\sim$PeV radiation by the 2.7K
cosmic microwave background.  The limits on the propagation distance
of Ultra High Energy (UHE) photons are so well established theoretically,
that possible detections$^{10,11}$ of two particular gamma-ray bursts
(GRBs) at, respectively, more than 50 TeV and more than 13 TeV, were
interpreted in the context of supposed Galactic origin of GRBs.

However, the {\sl observational}
 evidence for (or against) attenuation of the UHE
signal over large distances is less secure. It is true, that only
a few extragalactic sources have been reported, with the two
best established (Markarian 421 and 501) at a relatively close distance
(redshift 0.031 and 0.033, respectively), while brighter but more distance
blazars in the EGRET catalog (of sources of multi--GeV radiation) have
not been observed---this would be consistent with the predicted attenuation.
On the other hand, the fairly large
distance to these sources and the high energy range
of observation (above 20 TeV) were an embarassment to some fairly recent
estimates of the attenuation, so now the trend seems to be to
constrain$^{12,13}$
the IR by fitting the data on Mkn 421 and Mkn 501 to theoretical models
of UHE emission, and even to use the suggested attenuation to model
Galaxy formation.$^{14}$ Further, the observed power-law spectrum of Mkn 421
is clearly different from the ``curved'' spectrum of Mkn 501,
in spite of their nearly identical distance, strongly
suggesting$^{15}$ different
intrinsic spectra in different sources. Moreover, Mkn 421 is
known\footnote{With one flare so short (several minutes)
that it was possible to limit$^{17}$ the ``quantum gravity'' scale to
$E_Q\ge 4\times10^{16}\,$GeV, on the assumption that eq. (1) holds,
so that a dispersion would be expected$^3$ between higher and lower energy
photons from Mkn 421,
as discussed in Section 1.} to be
extremely variable$^{16}$.

 Thus, it is as yet uncertain whether the more distant AGNs are truly
invisible in UHE (e.g. 30 TeV) photons.  The same is even more true of
gamma-ray bursts, where models$^{18}$ of UHE emission and TeV
observations$^{10,11,19,20}$ are even less clear. However, as all GRBs
are thought to be cosmological in origin, with redshifts $z=0.835$,
$z=3.4$, and $z=0.97$ reported for three particular
sources$^{21,22,23}$, the importance of potential secure detections
of UHE photons in spatial and temporal coincidence with a GRB cannot
be overstated, as they could clearly indicate that the universe is
transparent to high energy photons.

  \section{Kinematics of pair creation}

It remains to show that with dispersion relation of eq. (1), pair creation
is forbidden for very asymmetric photons. This is related to the excess
momentum of the higher energy photon. Regardless, of whether an additional
term appears, or not, in the dispersion relation for the electron, as in
$$m^2c^4 +p^2c^2=E^2 +E^3/E_{QG}. \eqno (2)$$
and whether or not a term of the same order as the last term in eq. (2)
appears in the law of conservation of energy-momentum the result is the same
qualitatively: in addition to the usual threshold for pair creation,
$E_1E_2\ge m^2$, there is a maximum energy of the hard photon
$E_2\sim\sqrt{E_1/E_Q}$. This result is really a consequence of the
non-existence of the center of momentum when $E_2$ exceeds
$\sim2\sqrt{E_1/E_Q}$.

As a specific example, suppose that the ususal conservation law
is true
$$E_T=E_1+E_2=E_1'+E_2'\,,\ {\rm and}\ \bf p_1+p_2=p_1'+p_2'\,.\eqno (3)$$
Of course, all calculations must be done in a single frame of reference,
because we do not know how to transform frames.

Now, if eq. (2) holds (with eq. [1] a special case for m=0), and if two
particles of equal energy are created in a head-on collision of two
photons, then$^4$ eq. (3) holds, i.e., high energy photons are absorbed,
only if 
$$E_2\le 2\sqrt{2E_1/E_{QG}}. \eqno (4)$$

To illustrate the independence of the result on the dispersion relation,
and indeed the mass, of the electron, consider now that eqs. (1) and (3)
hold together with $m^2c^4 +p^2c^2=E^2$. Then pair creation by two
photons of energies $E_1$, $E_2$, is possible iff
$$E_1 E_2\ge m^2 c^4 +(E_1-E_2)^2(E_1+E_2)/(4E_{QG}),$$
yielding the usual threshold for symmetric energies, but also an upper limit
to the photon energy, similar to that of eq. (4)
$${m^2c^4\over E_1}\le E_2\le 2\sqrt{E_1E_{QG}}.$$

 \section{Conclusions}
If Lorentz invariance is modified, the kinematics of photon-photon collisions
(as well as of other transformations of particle identity) will certainly
be drastically modified. Although, there is no known self-consistent
theory of quantum gravity, if it is true that the dispersion relation of
 eq. (1) will hold for photons in a future theory, it seems very likely,
as for the specific examples above, e.g. eq. (4), that the universe will be
transparent to photons of energies higher than
$(30\,{\rm  TeV})\times\sqrt{E_Q/10^{17}\,{\rm GeV}}$.

  \end{document}